\documentclass[amsmath, amsfonts, twocolumn, prl]{revtex4}
\usepackage{graphicx}
\usepackage{epsfig}
\usepackage{bm}
\usepackage{euscript}
\usepackage{dcolumn}
\usepackage{amsmath}
\usepackage{amssymb}

\begin{document}

\title{Measurement of Coulomb drag between Anderson insulators}
\author{K. Elsayad}
\author{J. P. Carini}
\email[]{jpcarini@indiana.edu}
\author{D. V. Baxter}
\affiliation{Department of Physics, Indiana University, Bloomington, Indiana 47405} %

\begin{abstract}
We report observations of the Coulomb drag effect between two
effectively 2-d insulating a-Si$_{1-x}$Nb$_{x}$ films. We find that
there only exist a limited range of experimental parameters over
which we can measure a sizable linear-response transresistivity
($\rho_{\text{d}}$). The temperature dependence of $\rho_{\text{d}}$
is consistent with the layers being Efros-Shklovskii Anderson
insulators provided that a 3-d density of states and a localization
length smaller than that obtained from the DC layer-conductivity are
assumed.
\end{abstract}

\date{February 26, 2008}

 \maketitle


Materials such as a-Si$_{1-x}$Nb$_{x}$ which exhibit a disorder
driven $T=0$ (Quantum Critical) Metal-Insulator Transition (MIT)
 \cite{sgc97,bsd85lcb98} have presented many challenges to
condensed matter physics: in particular understanding the role of
long ranged electron-electron interactions in the insulating phases
\cite{hgc01} and in the vicinity of the MIT \cite{bsd85lcb98}. Since
the interplay between disorder and electron-electron interactions in
such systems will determine the dominant transport mechanism, the
development of experimental techniques to separately measure these
is useful. In this letter we show that the Coulomb drag effect
allows us to directly study long ranged electron-electron
interactions in insulating a-Si$_{1-x}$Nb$_{x}$ thin films. We find
that although linear-response Coulomb drag is only observable over a
limited range of sample parameters, when obtainable, it offers
unequivocal distinction to be made between alternative models for
the electronic transport in such systems.

Coulomb drag \cite{spf89} arises from the Coulomb scattering of
charge carriers in spatially separated layers, in the absence of
charge transfer between the layers. Experimentally, the Coulomb drag
effect between two layers (layer-$1$ and $2$) can be observed by
measuring the electric-field ($E_2$) created in one, open circuited,
layer due to a parallel applied current-density ($j_1$) in the
other. The (longitudinal) transresistivity, or linear-response
Coulomb drag coefficient, is defined as
$\rho_{\text{d}}=-E_{2}/j_{1}$; whilst the measured
total-transresistance is the ratio of the induced voltage in layer-2
to the applied current in layer-1, i.e. $-V_{2}/I_{1}$. Theoretical
analyses agree that the linear-response transresistivity between two
identical 2-d layers is given by, e.g. \cite{zm93}:
\begin{equation}
\label{eqn:1} \rho_{\text{d}} \sim \frac{\hbar^{2} \beta}{2 e^{2}
n^{2}} \int d\omega \int \frac{dq}{(2 \pi )^{2}}q^{3} \left| \frac{
\text{Im} \chi(\omega,q) U(\omega,q)}{\text{sinh}(\hbar\omega \beta
/2)} \right|^{2}
\end{equation}
where $\beta=(k_{\text{B}}T)^{-1}$, $T$ is the temperature, $n$ is
the carrier density in a layer, $\chi(\omega,q)$ is the
density-density response function of a layer, and $U(\omega,q)$ is
the screened interaction potential between the layers.
$\text{Im}\chi(\omega,q)$ may be obtained from the finite wavevector
conductivity via:
\begin{equation}
\label{eqn:flucdis} \text{Im}\chi(\omega,q) =
q^{2}(e^{2}\omega)^{-1}\sigma(\omega,q).
\end{equation}
The motivation for this study comes from predictions \cite{s97} that
the temperature dependence of the transresistance can be used to
differentiate insulating states. In particular, for the case of
(2-d) Mott Anderson insulator bilayers the low-$T$ transresistivity
should vary as $\rho_{\text{d}} \propto T^{2}$, whilst for (2-d)
Efros-Shklovskii(ES) Anderson insulators it should diverge when
$T\rightarrow 0$ as $\rho_{\text{d}}\propto
T^{3}\text{exp}[(T_{0}/T)^{1/2}]$, where $T_{0}$ is the ES
characteristic temperature given by $k_{\text{B}}T_{0}\approx
2e^{2}/\kappa\xi$ and $\kappa$ is the dielectric constant of the
layers. This \emph{opposite} behavior of the transresistivity as
$T\rightarrow0$ (diverging for ES-Anderson insulator bilayers, and
vanishing for Mott Anderson insulator bilayers) allows for a more
transparent distinction between the two insulating states over a
narrow $T$ range than intra-layer transport measurements do---where
one would obtain different functional $T$ or $\omega$ dependencies
with the same trend.

\begin{figure}
\includegraphics[width=7.5cm]{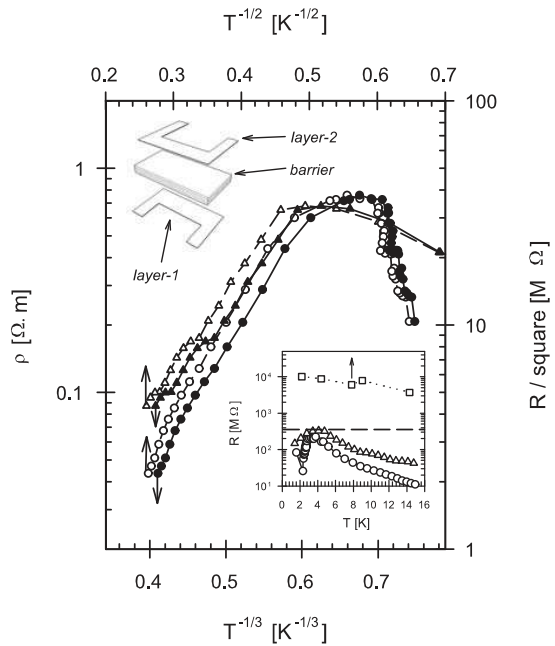}
\caption{\label{fig:2} Resistance per square of layers~$1$
(triangles) and $2$ (circles) in sample 1B. Open symbols are plotted
against the $T^{-1/2}$ axis and closed symbols against the
$T^{-1/3}$ axis. Upper inset: sketch of sample geometry. Lower
inset: resistances of layers-1 and -2 (triangles and circles)
compared to lower bounds of the barrier resistance (squares). The
dashed horizontal line represents the maximum resistance that could
be measured using $V_{\text{induced}}(I_{\text{applied}})$
techniques with existing apparatus.}
\end{figure}

In our study, samples consisted of two U-shaped 200\AA~thick
insulating a-Si$_{1-x}$Nb$_{x}$ layers, separated by a SiO barrier
(see upper left inset in figure~\ref{fig:2}). All layers were
fabricated using standard RF magnetron sputtering techniques, with
the a-Si$_{1-x}$Nb$_{x}$ layers being deposited using the
co-sputtering technique with a rotating sample-holder outlined in
Ref\cite{hgc01}. Samples were all grown on polished glass slides in
an inert (argon) environment at ambient temperature and a pressure
of $\approx0.50$kPa. Prior to the layers being deposited,
$600$\AA-thick silver connections and a silver bridge, that would
connect the two layers, were sputtered. The latter would prevent the
barrier from breaking due to electrostatic discharge/breakdown
during handling, characterization and cooling. Connections to the
signal generating and measuring apparatus were made with sputtered
silver layers on the arms of the U-shaped layers several millimeters
from the interacting-region. To prevent oxidation the top layer was
covered with a $\approx$50\AA-thick SiO film prior to atmospheric
exposure. Thicknesses were inferred from the sputtering rate
(measured using a quartz crystal thickness monitor), and were within
$\approx5\%$ of those obtained from optical interferometry
measurements on samples of similar compositions and thicknesses.
Estimates of the uncertainty in the thickness due to fluctuations in
the deposition rate and uncertainties in the exposure time are
$<\pm10$\AA~ and $<\pm15$\AA~ for the case of the
a-Si$_{1-x}$Nb$_{x}$ and the barrier layers respectively. The
barrier layer was deposited in discrete stages to reduce the
formation of pinholes. High sputtering powers and brief atmospheric
exposure between stages were found to increase the barrier strength
and durability. We were thereby able to fabricate barrier layers
with resistances several orders of magnitude larger than those of
the layers.

Samples were cooled to $T\approx1.2$K using a standard liquid helium
cryostat. The low-$T$ ($\leq$20K) resistance of the barrier layer
was obtained by measuring its conductance, i.e. by applying a small
DC potential difference between terminals on either layer, and then
measuring the induced current. Measurements were performed from
several terminals on each layer and with various terminals and
connections grounded to rule out any effects from ground-leakage and
-loops. Single layer resistance measurements were performed using
standard 4-wire techniques, for which it was assured that driving
currents were small enough to give linear-response coefficients. The
indifference of the layer resistances to the grounding of terminals
in the second layer provided further verification that inter-layer
leakage and tunneling was not significant. Transresistance
measurements were performed using a quasi-AC technique in which a DC
current source was programmed to flip polarity every $\sim$1 second.
Since in linear-response theory $\rho_{\text{d}}$ remains the same
on interchanging the layers (even if the layers have different
resistivities), the condition that
($R_{\text{drag}}$/square)$_{2\rightarrow 1}$ =
($R_{\text{drag}}$/square)$_{1\rightarrow2}$ as the driving-current
is decreased, was used as a test for the linear-response regime.
\begin{figure}
\includegraphics[width=8.5cm]{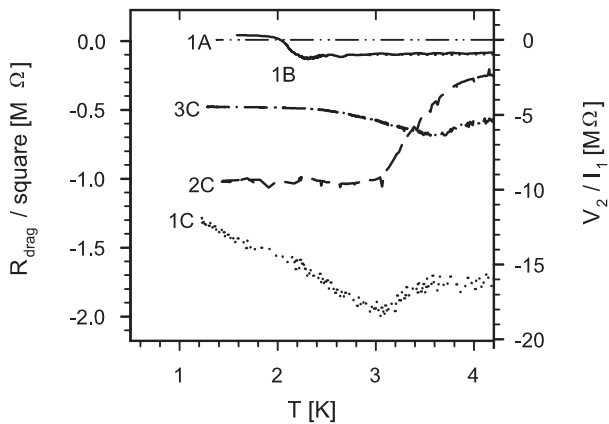}
\caption{\label{fig:1} Plot of the low-$T$ transresistance per
square ($R_{\text{drag}}$/square) for various samples, at a
driving-current of $I_1 = 1$nA. (The right-vertical axis shows the
ratio of the total measured voltage to the total driving current).
Samples 1A, 1B and 1C have an average Nb concentration of $x=0.070$
and layer separations of $200$\AA, $100$\AA~and $55$\AA~
respectively. Samples 2C and 3C have the same geometry as sample 1C
but with $x\approx0.076$ and $\approx0.080$.}
\end{figure}
The transresistance at driving currents of $I=1$nA for samples with
average Nb concentrations per layer \cite{niobiumconc} of
$x=0.07\rightarrow0.08$ and layer separations of
$50\rightarrow200$\AA~were found to, on average, increase with
decreasing temperature between $T\sim20\rightarrow3$K. At lower
temperatures the transresistance would saturate or decrease in
magnitude. This is shown in figure~\ref{fig:1} for a selection of
samples. In what follows we will present data for sample 1B, where
the transresistance entered the linear-response regime for
$I\leq1$nA. Due to the multitude of competing size effects (see
below), it was unclear, based on the temperature dependence of the
DC resistance alone, whether the layers are better described by the
ES Variable Range Hopping(VRH) model~\cite{es75}, which predicts:
$\text{ln}(\sigma_{dc})\propto T^{-1/2}$, or the Mott VRH
model~\cite{m70}, which, for effectively 2-d films, predicts:
$\text{ln}(\sigma_{dc})\propto T^{-1/3}$, at low temperatures. This
can be seen in figure~\ref{fig:2}, where we present $T^{-1/2}$ and
$T^{-1/3}$ plots of the layer-resistances of sample 1B. In the
temperature range $4\text{K}<T<15$K, ES VRH with a characteristic
temperature of $T_{0}=145(\pm\leq9)$K, 2-d Mott VRH with a
characteristic temperature of $T_{0}=1890(\pm60)$K, and 3-d Mott VRH
with a characteristic temperature of $T_{0}=36,000(\pm1000)$K, all
describe the $T$-dependence reasonably well. The localization length
determined from the ES characteristic temperature is
$\xi\approx105(\pm6)$\AA.
For each case the DC VRH transport in the layers should be
effectively 2-d for T$\leq$20K---on account of the resonant hopping
distance ($r_c$) being larger than the width of each layer ($W$). At
these temperatures the resistance of the barrier---obtained by
measuring the tunneling current (see above)---is found to be
approximately two orders of magnitude larger than the resistance of
layers-1 and -2 (see lower inset in figure~\ref{fig:2}).

The observed decrease of the layer-resistance and transresistance
observed at T$\leq$2.5K is (based on the tests outlined above and in
Ref\cite{gem91}) neither due to grounding loops/leaks or
tunneling/leakage through the barrier. It is unlikely that they are
due to thermoelectric effects (as in e.g. \cite{spf89}), given that
all connections were far away from the interacting region.
Furthermore, it is unlikely to be due to the layers undergoing a
superconductor-insulator transition---which is known to occur in
thin a-Si$_{1-x}$Nb$_{x}$ films with large Nb concentrations
($x\approx0.15-0.18$ \cite{amp06})---based on the high resistance
per square of the layers and the high temperature at which the
transition appears to occur. We will present a comprehensive study
of this phenomenon elsewhere, and focus on the temperature range
$2.5\text{K}<T<15$K in this letter.

\begin{figure}
\includegraphics[width=7.5cm]{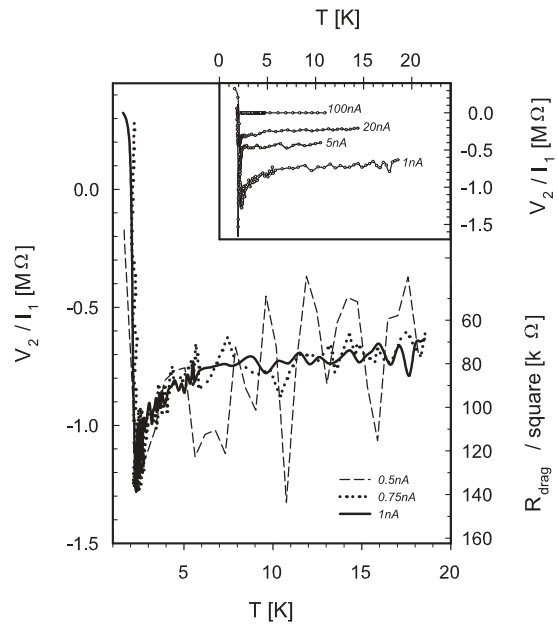}
\caption{\label{fig:3} Temperature dependence of linear-response
transresistance observed at driving currents $I\leq1$nA for sample
1B. Inset: At driving currents $I\geq1$nA. The anomalous low-$T$
increase of the $I<1$nA transresistance is consistent with the
layers being Efros-Shklovskii Anderson insulators and not Mott
Anderson insulators}
\end{figure}

As the driving current is increased the transresistance decreases
(see inset in figure~\ref{fig:3}), and for $I\approx100$nA, the
transresistance is found to be more than $3$ orders of magnitude
smaller than at $I=1$nA. This is likely due to larger driving
currents both increasing the effective sample temperature and
producing significant non-linear responses. For $I<1$nA the drag
voltage becomes noisier, but its average value changes by less than
$\approx5\%$ between $1$, $0.75$ and $0.5$nA (see
figure~\ref{fig:3}). We thus believe that at this point we have
reached the linear-response regime. It is unlikely that phonon
\cite{gem93} and plasmon \cite{fh94} contributions are significant
at these temperatures on account of the small layer separations and
the strong insulating nature of the layers.

\begin{figure}
\includegraphics[width=8.0cm]{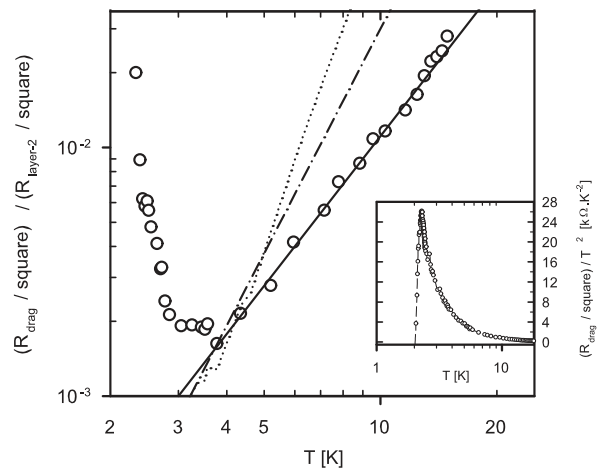}
\caption{\label{fig:4} Plot of the $T$-dependence of the ratio
($R_{\text{drag}}$/square) / ($R_{\text{layer-2}}$/square) at
$I=1$nA for sample 1B (circles), the predicted slope for the case of
two 2-d Mott-Anderson insulator layers (dotted line), the predicted
slope for the case of two 2-d Efros-Shklovskii Anderson insulator
layers (dash-dotted line), and the predicted slope for two
effectively 3-d Efros-Shklovskii Anderson insulator layers (solid
line). As can be seen, the last prediction describes the data best.
Inset: $T^2$-scaled transresistance per square for the same sample.}
\end{figure}

In order to limit the number of fitting parameters in analysing
these data, we find it convenient to plot the ratio
($R_{\text{drag}}$/square)/($R_{\text{layer}}$/square) as a function
of temperature. Doing so we find that our data is best described
(see figure~\ref{fig:4}) by the 2-parameter equation:
($R_{\text{drag}}$/square)$= aT^{b}(R_{\text{layer}}$/square), with
$a=1.1(\pm 0.1)\times 10^{-4}$K$^{-b}$ and $b=2.0(\pm 0.1)$.
The latter parameter deviates from the predictions of Ref\cite{s97},
which suggest that $b=3$ for a bilayer system comprised of 2-d ES
Anderson insulators.

The observed discrepancy can be explained if the screening in the
layers is not 2-d in the studied regime---i.e. if the response is
dominated by a finite wave-vector $q>W^{-1}$ (where
$W\approx200$\AA) for which the layers will be effectively 3-d.
Since $\xi^{-1}>W^{-1}>r_{\text{c}}^{-1}=\xi^{-1}(T_{0}/T)^{-1/2}$,
at the temperatures of interest, this occurs at momentum transfers
($q>r_{\text{c}}^{-1}$) that dominate the transresistance (see
Ref\cite{s97}).

Repeating the calculations for the finite-$\omega,q$ conductivity in
Ref\cite{as94} for 3-d systems, we find it takes the asymptotic
forms:
\begin{equation}
\label{eqn:3dfiniteqlow} \sigma_{3\text{d}}(\omega, q \ll
r_{\omega}^{-1})\sim
\mathcal{C}_{1}(e^{2}/\hbar)(\omega/\omega_{0})r_{\omega}^{-1}
\end{equation}
\begin{equation}
\label{eqn:3dfiniteq}
\sigma_{3\text{d}}(\omega, r_{\omega}^{-1}\ll q \ll \xi^{-1})\sim
\mathcal{C}_{2}(e^{2}/\hbar)(\omega/\omega_{0})q^{-2}r_{\omega}^{-3}
 \end{equation}
where $r_{\omega}=\xi\text{ln}(\omega_{0}/\omega)$,
$\omega_{0}=k_{\text{B}}T/\hbar$, and $\mathcal{C}_{i}$, $i=1,2$ are
numerical constants of order unity. Substituting
equations~(\ref{eqn:3dfiniteqlow}) and (\ref{eqn:3dfiniteq}) into
(\ref{eqn:flucdis}) and then (\ref{eqn:1}), we find that the
dominant contribution to the transresistance, if the layers are
treated as effectively 3-d for $q>r_{\text{c}}^{-1}$ and we can
assume weak static screening, would change with $T$ as:
$\rho_{\text{d}}\propto T^{4} \text{exp}[(T_{0}/T)^{1/2}]$. The
observed $\rho_{\text{d}}\propto T^{2} \text{exp}[(T_{0}/T)^{1/2}]$
temperature dependence can however arise if one or both of the
following are the case:
\\ {\bf (1)}~The relevant localization length is much smaller than
that obtained from the $T$-dependence of the DC conductivity so
that:
\begin{equation}
\label{eqn:lowe} \xi (T_{0}/T)^{1/2} < d.
\end{equation}
In this way the $r_{\text{c}}^{-1}<q$ momentum transfers (which are
otherwise dominant) are suppressed, and the $q<r_{\text{c}}^{-1}$
contribution determines the low-$T$ transresistance. Substituting
equation~(\ref{eqn:3dfiniteq}) into (\ref{eqn:flucdis}) and
(\ref{eqn:1}), the transresistance would now change with temperature
as: $\rho_{\text{d}} \propto T^{2} \text{exp}[(T_{0}/T)^{1/2}]$.
\\ {\bf (2)}~Finite-$\omega$ transport is effectively 3-d due to
pair-arms ($r_{\omega}$) reaching into the barrier and the second
layer. From the strongly localized nature of electrons in the
barrier layer, the effective $\xi$ would be much smaller and
condition (\ref{eqn:lowe}) may be satisfied, even if the effective
layer separation also decreases significantly.

We note that since several of the relevant length scales are
comparable ($r_{\text{c}}\sim\xi\sim W\sim2d$), a small modification
of the effective values that these parameters take (due to e.g.
finite-size effects, correlated hopping or surface effects) could
cause a change between the $\xi>d$ and $\xi<d$ regime in the
temperature range probed, resulting in a different $T$-dependence
than that predicted. We also note that we do not observe the
expected transition from the $q<r_{\text{c}}^{-1}$ to the
$q>r_{\text{c}}^{-1}$ regime, as the temperature is increased.
However, if the apparent change in the $T$-dependence of the
transresistance (see figure~\ref{fig:4}) at T$\approx$4K occurs when
the $q>r_{\text{c}}^{-1}$ regime kicks in, then we predict that
$\xi\approx(\text{4K}\beta\kappa/e^2)d^2 \sim 3$\AA. This would give
a transresistance that goes as $\rho_{\text{d}}\propto
T^{2}\text{exp}[(T_{0}/T)^{1/2}]$ for $T>4$K, or
$q<r_{\text{c}}^{-1}$, as we have observed.

In conclusion, we have observed the Coulomb drag effect between two
$200$\AA~thick insulating a-Si$_{1-x}$Nb$_{x}$ films (with $0.07\leq
x\leq0.08$) separated by a $50-200$\AA~thick SiO based barrier. We
were able to retrieve accurate linear-response data for
$x=0.070(\pm0.002)\%$, and layer separations of $100(\pm10)$\AA. The
temperature dependence of the transresistance in such samples was
found to be in agreement with that predicted for ES Anderson
Insulator films, provided that the localization length in the
a-Si$_{1-x}$Nb$_{x}$ layers is smaller than that inferred from the
temperature dependence of the DC layer-resistances. Our study
suggests that whilst theoretically the Coulomb drag effect is a
useful technique for distinguishing the insulating states of thin
films, it is experimentally challenging due to the complex
dielectric properties of disordered thin films at low energies,
which result in non-linear inter- and intra- layer excitations
becoming dominant at practical temperatures and sample dimensions.
Experimental studies of the non-linear crosstalk regime between thin
insulating films, along with simulations of the non-linear
(current-dependent) transresistance in such systems, may prove to be
the most productive method of studying the detailed nature of the
observed excitations.

\begin{acknowledgments}
We would like to thank Prof.~E.~Shimshoni for fruitful discussions,
and D.~Sprinkle and M.~Hosek for help with the experiments.
\end{acknowledgments}



\begin{thebibliography}{99}
\bibitem{niobiumconc} Average niobium concentrations were estimated by comparison of the deposition rate and characteristic ES and Mott temperatures to that of a-Si$_{1-x}$Nb$_{x}$ films
    for which $x$ was determined from Electron microprobe analysis.
\end{thebibliography}
\end{document}